# Optimization study of a Z-type airflow cooling system of a lithium-ion battery pack

Santosh Argade[1], Ashoke De[1,2 a)]

[1]Department of Sustainable Energy Engineering, Indian Institute of Technology Kanpur, 208016, Kanpur, India.

[2]Department of Aerospace Engineering, Indian Institute of Technology Kanpur, 208016, Kanpur, India.

The present study aims to optimize the structural design of a Z-type flow lithium-ion battery pack with a forced air-cooling system (FACS) known as BTMS (Battery Thermal Management System). The main goal is to minimize $T_{max}$ (maximum temperature) and $\Delta T_{max}$ (maximum temperature difference) while ensuring an even airflow distribution within the battery module. The present study thoroughly investigates critical factors such as the inlet air velocity, tapered inlet manifold, and the number of secondary outlets to evaluate their impact on thermal performance and airflow uniformity within the battery module. Increasing the inlet air velocity from 3 to 4.5 m/s significantly improves the thermal cooling performance of the BTMS, resulting in a decrease of 4.57 °C (10.05%) in $T_{max}$ and 0.29 °C (9.79%) in $\Delta T_{max}$ compared to the original 3 m/s velocity. Further, the study assesses the significance of a tapered inlet manifold as a critical factor, revealing its substantial impact on cooling performance and temperature reductions in battery cells 3-9. It also facilitates a more uniform airflow distribution, decreasing the velocity difference between channel 9 and channel 1 from 3.32 m/s to 2.50 m/s. Incorporating 7 secondary outlets significantly improves the heat dissipation ability of the BTMS, resulting in a decrease of 0.894 °C (2.18%) in $T_{max}$ and 2.23 °C (72.84%) in $\Delta T_{max}$ compared to the configuration with 0 secondary outlets. By optimizing these parameters, the aim is to enhance BTMS's capabilities, improving LIB (Lithium-Ion Battery) packs' performance and reliability. The optimized structural design parameters proposed in this study yield practical applications that extend beyond theoretical insights, impacting diverse fields reliant on lithium-ion battery technology. Through enhanced thermal management systems, applications in electric vehicles and portable electronics are poised to experience improved performance and longevity. Furthermore, these advancements inform the development of next-generation battery packs, promising reduced overheating risks and extended battery life. Such innovations are critical in energy storage systems for renewable energy applications and electric vehicle technology, facilitating faster charging times and increased driving range. Moreover, the implications extend to aerospace applications, ensuring the reliability of batteries in extreme environmental conditions.

___________________________

a) Author to whom correspondence should be addressed. Electronic mail: ashoke@iitk.ac.in

**INTRODUCTION**

A crucial step towards improving the deteriorating climate conditions is the electrification of transportation systems. The urgent need for vehicle electrification to mitigate pollution and reduce reliance on fossil fuels has propelled the widespread adoption of lithium-ion batteries (LIBs) in electric vehicles (EVs). LIBs are crucial in developing EVs and providing efficient energy storage and delivery. LIBs have overtaken other competing battery chemistries due to their intrinsic qualities, including high power density, energy density, prolonged cycle life, and low self-discharge rate. However, if run outside the preferred thermal (i.e., temperature) and electrical (i.e., voltage and current) limits, these batteries experience several performance and safety problems. Additionally, the LIB produces a large quantity of heat while functioning, and the amount varies depending on the circumstances. The batteries are outfitted with the thermal management system, which guarantees battery operation within predetermined temperature ranges and controls these difficulties. A wide range of cooling techniques exists for batteries, such as heat pipe, air, liquid, and PCM cooling. With many merits, including a simple structure, a low price, minimal leaks, and no added weight, air cooling has emerged as a simple way to accomplish these objectives.

Liquid cooling systems excel in dissipating heat from battery modules due to their superior heat capacity and thermal conductivity compared to air. This leads to enhanced heat transfer efficiency and lower operating temperatures, prolonging battery lifespan. Moreover, liquid cooling systems boast compactness and lightweight design, which are ideal for applications with space constraints. Conversely, air-cooling systems offer cost-effectiveness and simplicity in implementation and maintenance. They require fewer components and infrastructure, providing greater scalability and adaptability to various vehicle architectures. However, a comprehensive analysis must consider factors such as reliability, safety, and environmental impact beyond efficiency and cost. Liquid cooling systems may face challenges like leakage and corrosion, jeopardizing system integrity, whereas air-cooling systems are less susceptible. Nevertheless, air-cooling systems may encounter limitations in extreme conditions or high ambient temperatures.

In contrast to previous works that have explored individual optimization strategies, this study investigates multiple structural optimization parameters. It delves into the effects of inlet air velocity, tapered inlet manifold, and the number of secondary outlets on thermal performance and airflow uniformity within the battery module. Therefore, while previous literature may have examined various aspects of battery cooling optimization separately, this study offers a comprehensive analysis by considering multiple optimization factors simultaneously.

According to earlier research, power batteries should have a $\Delta T_{max}$ of less than 5°C[1,2,3,4]. They can function normally when their $T_{max}$ (maximum temperature) is between 20°C and 45°C[1,2,3,4]. However, the operation of the battery involves internal chemical reactions that result in the generation of significant Joule heat. Without an efficient cooling system, the battery's temperature will rapidly rise with increased heat. The battery pack will be damaged and fail, and there may even be significant safety incidents if the battery temperature exceeds what is allowed. An effective thermal management system is essential for efficiently cooling the battery and ensuring optimal performance[5]. According to the Arrhenius law, a battery's performance is directly correlated with its temperature, as indicated by the Arrhenius law, which states that chemical reaction rates exponentially increase with temperature. The hottest region in a battery cell, according to research reports[6], is often found close to the electrode, resulting in temperature non-uniformity and reduced cell lifespan and performance. A battery's internal heat generation causes it to fluctuate in temperature. This internal heat generation is influenced by factors such as ohmic heat, mixing heat, enthalpy heating, and entropy, which directly impact the electrochemical reaction[7].

Measuring heat generation in the battery at different C-rate discharges is essential. The study conducted by Drake et al.[8] involved measuring temperature and heat flux to assess Li-ion cells' heat generation characteristics and thermal behavior during high C-rate discharges. The findings revealed a non-linear relationship between heat generation and C-rates, highlighting the significance of accurate measurement and analysis. A study by Srinivasan et al.[9] focused on measuring heat generation in a Li-ion cell during discharge, including electrolyte resistance ($R_s$), anode resistance ($R_a$), cathode resistance ($R_c$), and entropy variations in the cathode ($\Delta S_c$) and anode ($\Delta S_a$). The study introduces a method to detect heat production from each source separately, providing insights into the thermal behavior of the cell's components and emphasizing the importance of real-time monitoring for precise measurement and battery safety. Chen et al.[10] addressed the research topic of heat production in prismatic LFP batteries, focusing on a 20 Ah A123 LFP battery. The study reveals an inverse relationship between operational temperatures and heat generation rate, while higher discharge rates lead to increased heat generation.

A wide range of cooling techniques exists for batteries, such as heat pipe, air, liquid, and PCM cooling. Al-Zareer et al.[11] examined the improvements in BTMSs and their impact on battery performance, lifespan, and safety, including air, liquid, and PCM-based cooling systems. The study explores strategies such as maximizing air delivery, optimizing coolant selection and contact region, and introducing submerged batteries in compressed liquid coolant for enhanced thermal management. The study by Ling et al.[12] proposed a hybrid thermal management system for LIBs, combining FACS and PCMs to improve battery performance and safety. This strategy integrates active air circulation for heat dissipation and PCMs for efficient thermal buffering, resulting in excellent cooling performance and reduced thermal stress on battery cells.

Air cooling has become a popular and cost-effective method for cooling batteries due to its simple structure, affordability, minimal leaks, and absence of added weight. The first factor to consider in air cooling is the ventilation method. Pesaran et al.[13] used finite element analysis to compare cooling performance in parallel and series ventilation systems. The study found that the parallel ventilation system showed significant improvements with a 10 °C reduction in $T_{max}$ and a 4 °C decrease in $\Delta T_{max}$, outperforming the series ventilation system. Research by Na et al.[14] demonstrated that reverse-layered airflow, compared to unidirectional airflow, provides better battery temperature control. Using CFD simulations and experimental measurements, they found that incorporating rectifier grids further improved backward-layered airflow, reducing peak temperature by 0.5 °C and the average temperature by 2.7 °C (54.5%). Zhuang et al.[15] propose a novel BTMS incorporating intelligent cooling, reciprocating air flow, and structural optimization to enhance cooling efficiency and temperature uniformity. ANSYS simulations demonstrate a significant 76.4% energy savings and reduced temperature non-uniformity from 1.5 °C to approximately 0.6 °C compared to conventional systems. A numerical study by Chen et al.[16] suggests using a symmetrical BTMS with air cooling to better cool battery packs. The symmetrical system, with various inlet and outlet positions, reduces temperature variations ($\Delta T_{max}$) by at least 43% compared to asymmetrical systems, improving cooling efficiency and lowering energy consumption. Temperature variations in the battery cells are influenced by the arrangement of battery cells in the battery module. Fan et al.[17] studied three array configurations in battery packs and found that the aligned pack outperforms staggered and cross packs in cooling efficiency and temperature uniformity. Higher air inlet velocities improve cooling but negatively affect energy efficiency, while discharge rate affects cooling capacity. The type of cooling media used for cooling a battery module significantly impacts battery performance. Akbarzadeh et al.[18] compared the thermal performance of air and liquid cooling systems for a 48V battery module. The study found that the liquid cooling system effectively reduced the hottest cell's temperature and temperature differentials between cells, emphasizing the benefits of liquid cooling for efficient thermal management of battery modules.

Scholars have made numerous attempts and research to enhance the cooling performance of air-cooled BTMS by optimizing their structural parameters. Xie et al.[19] focused on optimizing FACS for heat dissipation in LIBs.

They used CFD models and experiments to find the best combination of inlet and outlet manifold angles and airflow channel widths, achieving a 12.82% reduction in $T_{max}$ and a 29.72% reduction in $\Delta T_{max}$, demonstrating the effectiveness of their approach for enhancing thermal performance in BTMSs. Zhang et al.[20] optimized air cooling in BTMS with CFD, introducing baffles and secondary outlets to reduce $T_{max}$ by 2.17 °C (4.95%) and $\Delta T_{max}$ by 91.89%. Wang et al.[21] proposed a method of attaching parallel plates to optimize airflow dispersion in air-cooled BTMS, resulting in a substantial 6.26% reduction in $T_{max}$ and a significant 90.78% decrease in $\Delta T_{max}$, improving cooling efficiency. Wu[22] conducted a study on enhancing the FACS of U-type BTMS in prismatic LIBs. The study analyzed battery spacing, tapered manifolds, and airflow rates to optimize cooling efficiency, achieving reduced power consumption and improved cooling capacity. Shi et al.[23] developed an improved U-type BTMS with 3 additional 4 mm airflow outputs to enhance temperature uniformity in LFP cuboid battery modules, achieving a 6.22K $T_{max}$ reduction and 40.36% $\Delta T_{max}$ improvement. CFD simulations validated their AI model with average absolute errors of 0.99% for $\Delta T_{max}$ and 0.046% for $T_{max}$. Xu et al.[24] conducted direct numerical simulations to explore the impact of wall shear on thermal convection. Their findings reveal that stronger wall shear enhances heat transfer efficiency and can lead to thermal turbulence relaminarization. This study provides insights into the dynamics of wall-sheared thermal convection, highlighting the trade-off between improved heat transfer and increased mechanical energy expenditure.

Focusing on the urgent need for vehicle electrification to mitigate pollution and reduce reliance on fossil fuels, the primary aim is to drive the widespread adoption of LIBs in EVs. While LIBs present various advantages over alternative battery technologies, they exhibit limitations, notably the excessive heat generation during discharge and charge cycles. Effective thermal management techniques are indispensable for ensuring peak battery performance and extended lifespan. The central goal involves optimizing the structural design of a Z-type flow LIB pack integrated with a FACS known as BTMS. The overarching objective is to minimize two key factors: $T_{max}$ (maximum temperature) and $\Delta T_{max}$ (maximum temperature difference) within the battery setup. Simultaneously, the even distribution of airflow within the battery module remains a crucial concern. In contrast to previous studies that typically investigate the impact of one or two parameters on the performance of BTMS, our study takes a novel approach by simultaneously exploring multiple parameters and their combined effects on BTMS. Aspects like the velocity of the inlet air, implementation of a tapered inlet manifold, and configuration of secondary outlets are subject to thorough investigation to assess their impacts on thermal efficiency and airflow uniformity within the battery module. Meticulous optimization of these vital parameters strives to elevate the thermal management capabilities of BTMS, effectively enhancing the performance and dependability of LIB packs. The findings derived from this study play a pivotal role in advancing the development of finely tuned cooling systems for LIBs, significantly contributing to the progression of efficient and sustainable energy storage solutions tailored to the unique demands of electric vehicles.

## I. NUMERICAL METHODOLOGY

### A. Governing Equations

The Navier-Stokes equations govern fluid flow and are based on Newton's second law of motion and the constitutive relationship involving viscous forces. These equations explain how momentum and mass are conserved during fluid flow. The NS equations apply to incompressible fluids under the Newtonian fluid assumption, where the relationship between stresses and deformation rates is linear. The equations comprise the momentum equation, which considers pressure, viscous, and external forces impacting the fluid flow, and the continuity equation, which ensures mass conservation, as explained in the section below.

#### a. Continuity Equation

$$\frac{\partial u_i}{\partial x_i} = 0 \tag{1}$$

Where $u_i$ is the velocity in the *x*, *y,* and *z* directions.

#### b. Momentum Equation

$$\rho_a u_j \frac{\partial u_i}{\partial x_j} = -\frac{\partial p}{\partial x_i} + \frac{\partial}{\partial x_j}\left[(\mu + \mu_t)\frac{\partial u_i}{\partial x_j}\right] \tag{2}$$

Where $u_i$ and $u_j$ are components of velocity, $\rho_a$ is the density of air, $p$ is Reynolds average pressure, and $\mu_t$ is turbulent viscosity coefficient.

#### c. Energy Equation

$$\rho_a C_{p,a}\frac{\partial T_a}{\partial t} + \rho_a C_{p,a} u_j \frac{\partial T_a}{\partial x_j} = \frac{\partial}{\partial x_j}\left[\left(\lambda_a + \frac{\mu_t}{\sigma_T}\right)\frac{\partial T_a}{\partial x_j}\right] \tag{3}$$

Eqs. (3) represents the temperature equation for the airflow region.

$$\rho_b C_{p,b}\frac{\partial T_b}{\partial t} = \nabla.(\lambda_b \nabla T_b) + q_{gen} \tag{4}$$

Eqs. (4) represents the temperature equation for the battery region.

Where the variables for the air and the battery are denoted by the subscripts "$a$" and "$b$". The temperature, thermal conductivity, heat capacity, and turbulence model parameters are denoted as $T$, $\lambda$, $C_p$ and $\sigma_T$ respectively. $q$ denotes the volumetric heat generation of LIB.

### d. Heat Generation of LIB

Two components comprise most of the heat the battery produces ($Q_{gen}$). One is the heat that the battery's internal resistance produces irreversibly. The second one is the reversible heat produced in the battery by the electrochemical reaction. The standard Bernadi heat generation model often describes the heat generation rate as follows[25]:

$$Q_{gen} = I\left[(U_{OCV} - U) + T_b \frac{\partial U_{OCV}}{\partial T_b}\right] \tag{5}$$

Were,

$I(U_{OCV} - U)$ : Joule heat produced by the battery's internal resistance (Irreversible).

$IT_b \frac{\partial U_{OCV}}{\partial T_b}$ : Heat produced by the electrochemical reaction inside the battery (Reversible).

As a result, the heat generation of batteries may also be expressed as shown in[26]:

$$Q_{gen} = I^2 R + IT_b \frac{\partial U_{OCV}}{\partial T_b} \tag{6}$$

The battery absorbs the following amount of heat as the temperature rises:

$$Q_{abs} = m_b C_{p,b} \frac{dT_b}{dt} \tag{7}$$

The heat generated by the battery is equal to the heat absorbed by the battery when it is placed in an insulated environment,

hence equating Eqs. 6 and 7,

$$m_b C_{p,b} \frac{dT_b}{dt} = I^2 R + IT_b \frac{\partial U_{OCV}}{\partial T_b} \tag{8}$$

$$\frac{1}{I}\frac{dT_b}{dt} = \frac{R}{m_b C_{p,b}} \cdot I + \frac{1}{m_b C_{p,b}} \cdot T_b \cdot \frac{\partial U_{OCV}}{\partial T_b} \tag{9}$$

Zhang et al.[27] conducted a study covering the battery with heat-insulated cotton to create an adiabatic atmosphere. They discharged the battery at five distinct rates (0.5C, 1C, 1.5C, 2C, and 2.5C), investigated the average temperature increase curve, and obtained five different $\frac{\partial T_b}{\partial t}$ values. Table 1 presents the parameters for the prismatic lithium iron phosphate (LFP) battery under investigation. Chacko et al.[28] and Panchal et al.[29] suggest that when the battery operates at 25 – 45 °C with SOC values of 0.2 – 0.9, the joule resistance R can be considered a constant. The term $\left[T_b \left(\frac{\partial U_{OCV}}{\partial T_b}\right)\right]$ describes the battery's internal electrochemical process, which remains constant for the same battery, relying on the SOC value and the battery's type[30]. The term $\frac{dT_b}{dt}$ represents a constant for constant-current discharge for 15 minutes[31]. We can consider the function $\frac{dT_b}{Idt}$ as having a linear relation with the current I in Eqs. 9.

The linear relationship between $\frac{dT_b}{Idt}$ and $I$ is as follows.

$$\frac{1}{I}\frac{dT_b}{dt} = 7 \times 10^{-6}I + 2 \times 10^{-4} \tag{10}$$

Eqs. 9 states that $\frac{R}{m_b C_{p,b}}$ represents the slope of the linear line. The calculations reveal that the battery's equivalent specific heat $(C_{p,b})$ is 1633 $J/(kg.K)$. By combining Eqs. 7 and 9, the heat generation of the battery $Q_{gen}(W)$ is expressed as follows:

$$Q_{gen} = m_b C_{p,b} \frac{dT_b}{dt} = 4 \times 10^{-3} I^2 + 0.114I \tag{11}$$

Therefore, we can express the heat generation of the battery per unit volume $\left(q_{gen}\right)$ or volumetric heat generation as follows:

$$q_{gen} = \frac{Q_{gen}}{V_b} = 24.42I^2 + 695.97I \tag{12}$$

Using Eqn. 11, Table 2 presents the calculation of heat generation by the battery at various discharge rates ranging from 0.5 C to 2.5 C.

Table 1. Parameters of a LIB[1]

| Parameter | Numerical value |
|---|---|
| Nominal capacity $(Ah)$ | 15 |
| Nominal voltage $(V)$ | 3.2 |
| Charge upper limit voltage $(V)$ | 3.65 |
| Discharging cut-off voltage $(V)$ | 2.25 |
| Charging current $(C)$ | 0.5 |
| Resistance $(m\Omega)$ | 4 |
| Cell mass $(kg)$ | 0.35 |
| Cell dimensions $(mm)$ | 140 × 65 × 18 |

Table 2. At different discharge rates heat generation by the battery $Q_{gen}$

| Discharge Rate ($C$) | Heat generated by battery ($W$) |
|---|---|
| 0.5 | 1.08 |
| 1 | 2.61 |
| 1.5 | 4.59 |
| 2 | 7.02 |
| 2.5 | 9.90 |

Utilizes the Open FOAM chtMultiRegionSimpleFoam solver to model heat transfer across multiple regions. Based on the Finite Volume Method (FVM), the solver can handle steady-state or transient convection-diffusion equations with different boundary conditions and thermal characteristics. It employs the SIMPLE algorithm to establish a connection between pressure and velocity fields, ensuring mass and energy conservation. Sets the convergence residuals of the governing equations to 10$^{-6}$ and utilizes a second-order upwind discretization scheme for momentum, energy, and pressure.

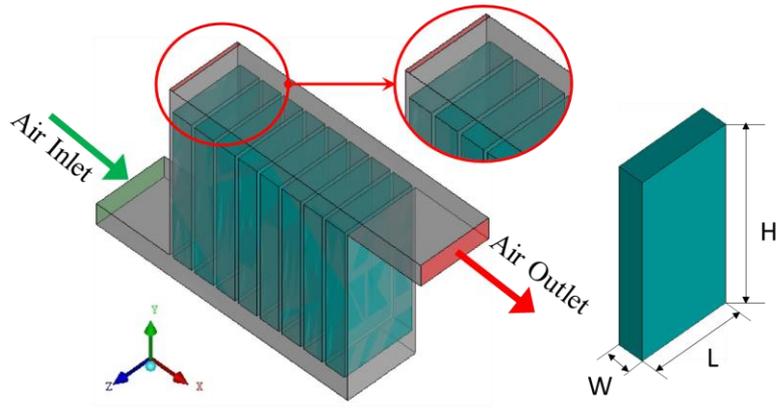

(a)

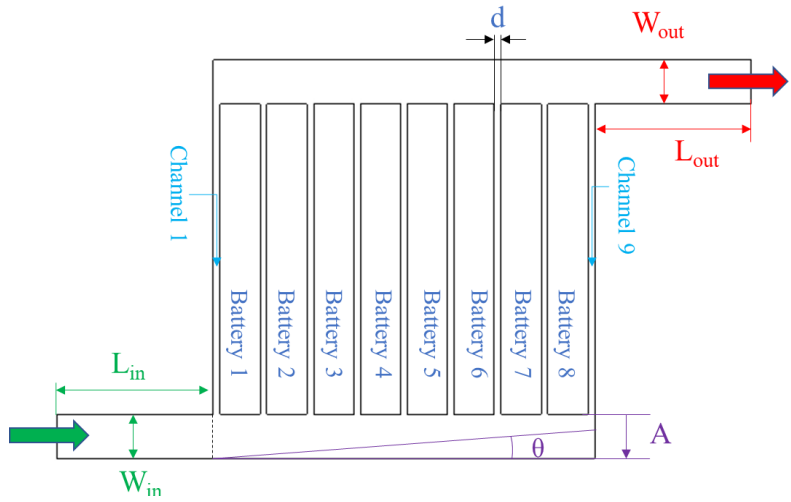

(b)

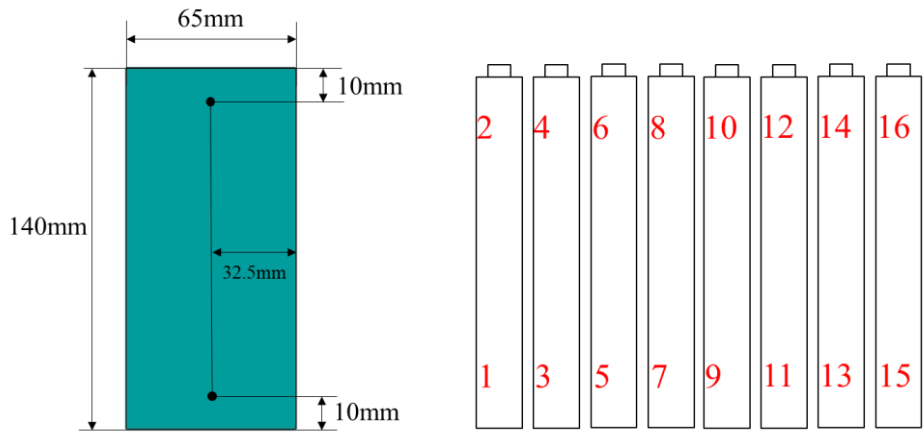

(c)

FIG. 1: (a) 3D Z-type parallel air-cooled BTMS schematic representation (b) Detailed representation of the dimensional parameters of BTMS, illustrating the specific measurements and geometric characteristics essential for performance evaluation (c) Location of the temperature measurement points

## II. MODEL DESCRIPTION AND COMPUTATIONAL DETAILS

### A. Z-Type Parallel Air Cooled BTMS

The study utilizes a Z-type parallel air cooling BTMS, depicted in Fig. 1 (a), with the air inlet and outlet on opposite sides of the battery module. The BTMS incorporates an additional secondary outlet positioned above channel 1, matching the cooling channel's width of 3mm and extending the length of 65mm, corresponding to the cooling channel and battery cell.

Fig. 1 (b) displays the specifications of the battery pack model, which consists of 8 battery cells, 9 airflow cooling channels, and battery cells connected to both sides of the battery box. The battery cell dimensions are as follows: length of 65 mm, width of 18 mm, and height of 140 mm. Each cooling channel has a width (d) of 3 mm, inlet and outlet width (Win and Wout), and inlet and outlet length (Lin and Lout) of 20 mm and 70 mm, respectively. The inlet manifold angle (A) is measured from the bottom of the battery cell.

### B. Computational Domain Details and Boundary Conditions

Furen Zhang et al.[20] experimented by substituting the battery cell with an aluminum block and heating rods. They connected four heating rods in parallel within each aluminum block to a DC power supply. This setup generated heat equivalent to the battery's discharge rate, specifically 60439.56 W.m−3 for a 2.5C discharge rate. They affixed two thermocouples to the surface of each battery, as depicted in Fig. 1 (c), to measure and collect temperature data. The battery's temperature is determined by averaging the readings from the two thermocouple points. In total, 16 thermocouples are placed on 8 batteries, as shown in Fig. 1 (c).

Furen Zhang et al. [20] performed an experimental study with specific initial conditions. The experiments were conducted at an ambient temperature of 25 °C, a discharge rate of 2.5C, and two different inlet air flow rates of 3 and 3.5 m/s. In their experiment, air and battery cells were considered the fluid and solid domains, respectively. In contrast, an aluminum block and heating rod replaced the battery cell in an experiment[20]. The battery material was assumed to be aluminum for simulation verification, but the battery cell material was chosen as lithium in the subsequent optimization simulations. The material properties of the air and battery used in the simulations are provided in Table 3.

Table 3. Properties of air and battery cell[1]

| Property | Air | Battery (Lithium) | Battery (Aluminum) |
|---|---|---|---|
| Density ($kg.m^{-3}$) | 1.165 | 2136.8 | 2700 |
| Specific heat ($J.kg^{-1}.K^{-1}$) | 1005 | 1633 | 900 |
| Dynamic viscosity ($kg.m^{-1}.s^{-1}$) | $1.86 \times 10^{-5}$ | - | - |
| Thermal conductivity ($W.m^{-1}.K^{-1}$) | 0.0267 | 1 (in the thickness direction) 29 (along surfaces) | 240 |
| Volume heat source ($W.m^{-3}$) | - | 60439.56 | 60439.56 |
| Size (($H$)$mm \times$ ($W$)$mm \times$ ($L$)$mm$) | - | 18 × 65 × 140 | 18 × 65 × 140 |

The inlet velocity is defined as the inlet condition, while the outlet condition is defined as the standard atmospheric outlet. The ambient temperature is assumed to be equal to the inlet air temperature. The walls are treated as adiabatic non-slip walls using the wall function approach and the SIMPLE algorithm. The momentum, energy, and pressure equations are discretized using a second-order upwind scheme, and the convergence residuals for the governing equations are set to 10$^{-6}$. The simulations are performed using the open-source software OpenFOAM to solve and simulate the governing equations for the air-cooled system.

Simulating temperature and airflow fields in a Z-type air-cooled BTMS involves using a CFD approach to consider the actual airflow and heat generation inside the battery. Several assumptions simplify the computations and effectively model the complex process.

i. Since the cooling airflow had a Mach number (*Ma*) ≪ 0.3, the air is considered an incompressible fluid.
ii. The effect of air buoyancy is ignored.
iii. The contact surfaces are coupled, and there is no relative slip between the fluid and the solid.
iv. The battery's density and specific heat capacity remain constant, and the physical properties of the different materials within the battery are uniform.
v. The thermal conductivity of a battery is anisotropic.
vi. The negligible impact of internal battery radiation on heat dissipation allowed it to be ignored.
vii. During charging and discharging, the battery's current density is evenly distributed, and each component generates heat in an equal amount.

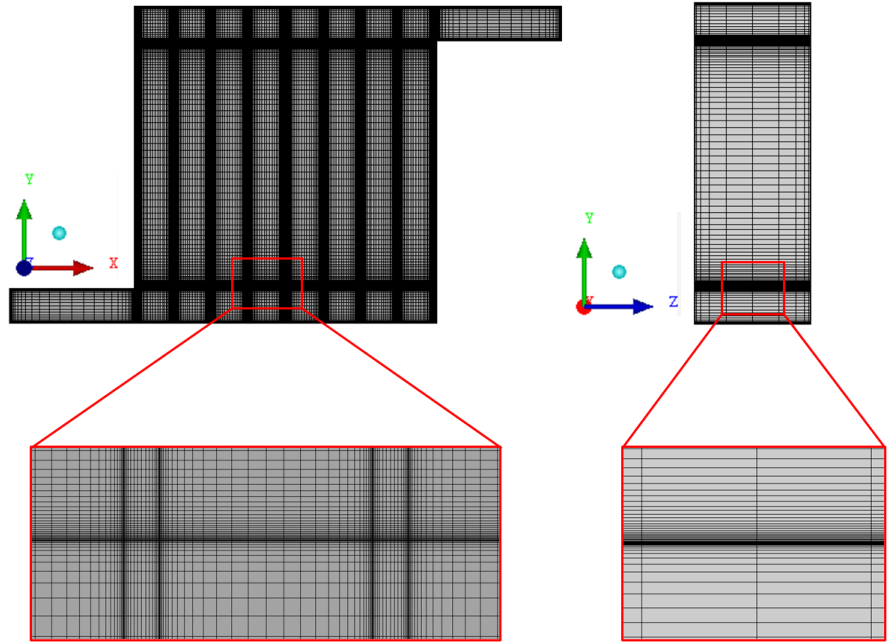

FIG. 2: Illustration of the structured computational grid used for the simulation, showing the refinement in the boundary layer region to capture detailed flow characteristics

## C. Grid Details

The presented study employs the CFD method to compute the velocity field and temperature distribution of the BTMS. The flow in the system is turbulent and modeled using the k-w SST (Shear Stress Transport) model. This choice is made due to the advantages the k-w SST model offers over the standard k-ε and k-w models, such as its improved accuracy in predicting complex turbulent flows and its ability to handle adverse pressure gradients more effectively. The multi-block structured grid is generated using ICEM CFD software for the simulation. To ensure accurate turbulence modeling, the first cell height on the wall is adjusted to maintain a Y+ value close to 1. The Reynolds number, calculated based on the inlet air velocity, exceeds 2300, confirming the presence of turbulent flow conditions within the BTMS. The initial layer grid height is 0.1 mm for enhanced simulation accuracy. This grid configuration, combined with a cell spacing progression ratio of 1.1 normal to the wall, ensures consistent grid spacing throughout the computational domain.

Multiple blocks are generated to manage the grid and ensure effective proper resolution. Fig. 2 provides a visual representation of the generated structured grid, highlighting the boundary layer grid and the overall grid structure of the cooling BTMS. This grid configuration is designed to accurately capture the flow characteristics and boundary layer behavior during the simulation.

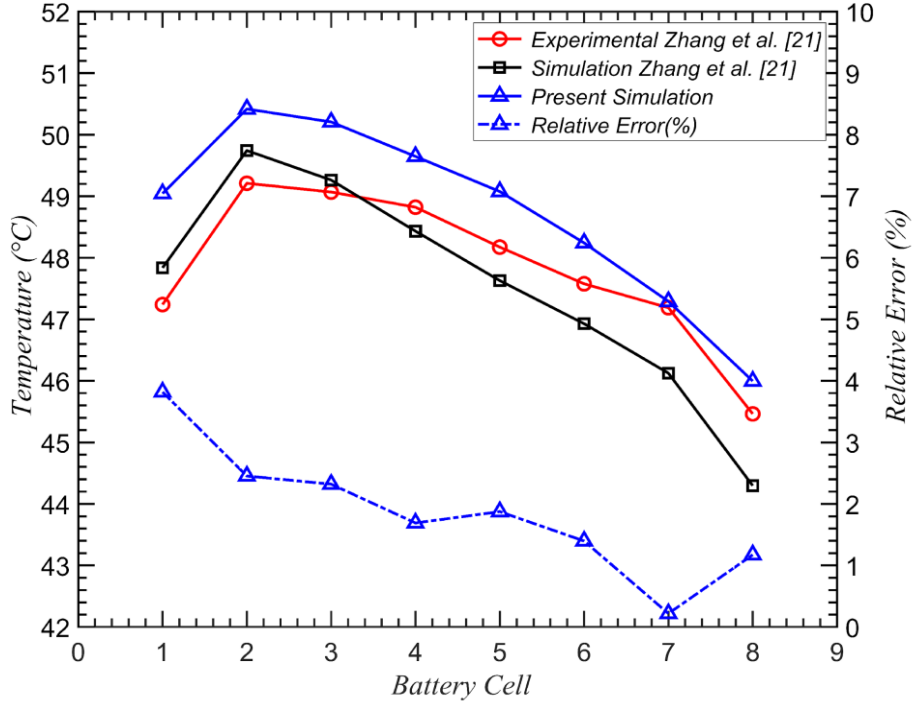

FIG. 3: Comparison of battery temperature profiles between experimental measurements and simulated results at an airflow rate of 3 m/s

## III. MODEL VALIDATION AND GRID INDEPENDENCE

### A. Model Validation

We have initially compared our temperature predictions of each battery with the experimental data of the literature[20]. The average temperature of each battery is determined by taking measurements from the thermocouples placed on the battery surface, as depicted in Fig. 3.

The comparison of the experimental and simulated battery cell temperatures and the resulting simulation findings are presented in this study. These results are displayed in Fig. 3 and Fig. 4. A 25 °C ambient temperature, a 2.5C discharge rate and inlet air velocities of 3 & 3.5 m/s are used as the operational parameters. The findings show a similar trend between the experimental and simulated battery temperatures. The comparison reveals errors within 3.8% for each battery temperature at an inlet velocity of 3 m/s, while at an inlet velocity of 3.5 m/s, the errors are within 4.9%.

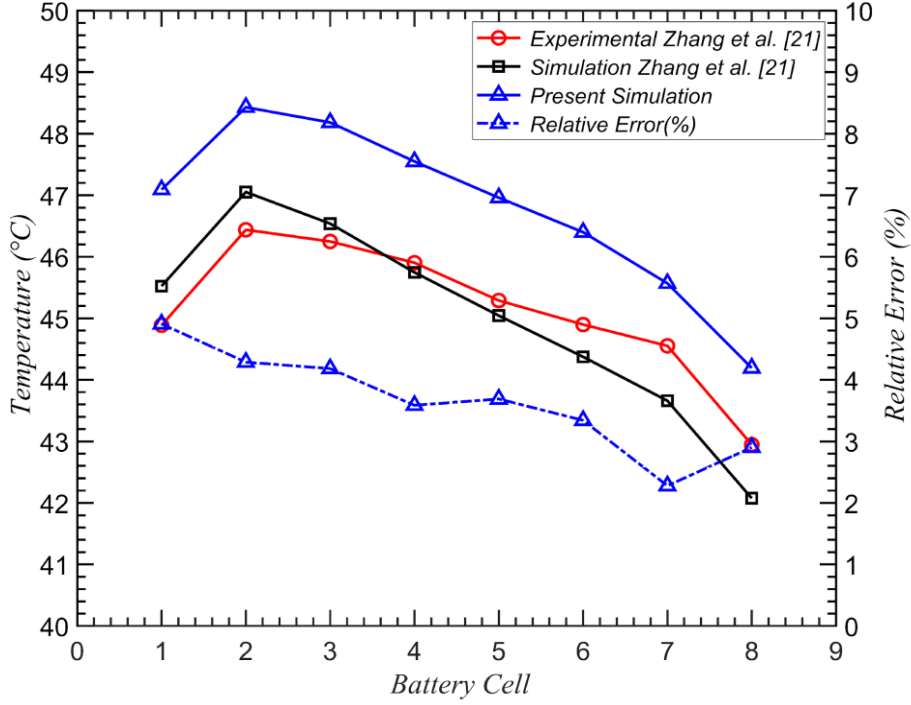

FIG. 4: Comparison of battery temperature profiles between experimental measurements and simulated results at an airflow rate of 3.5 m/s

These findings suggest the accuracy and effectiveness of the CFD method and model used in this study for predicting battery temperatures. The excellent concordance between the experimental and simulated results validates the simulation model's validity.

**B. Grid Independence**

Lithium is chosen as the battery material for further optimization, and the operational parameters are set to an ambient temperature of 25 °C, inlet air velocities of 3 m/s, and a discharge rate of 2.5C. A grid independence study determines the appropriate grid size for the battery package model in a CFD simulation, as the number of grids affects the simulation results. The $T_{max}$ and $\Delta T_{max}$ of the BTMS are examined under various numbers of grids, as shown in Fig. 5.

$T_{max}$ and $\Delta T_{max}$ change relatively little as the number of grids increases. When the number of grids exceeds $1.6 \times 10^6$, $T_{max}$ and $\Delta T_{max}$ remain unchanged, with less than 0.01 °C changes. Therefore, a grid number of $1.6 \times 10^6$ is selected for creating the meshes in all the models used in this study. This choice ensures the necessary accuracy while reducing computation time.

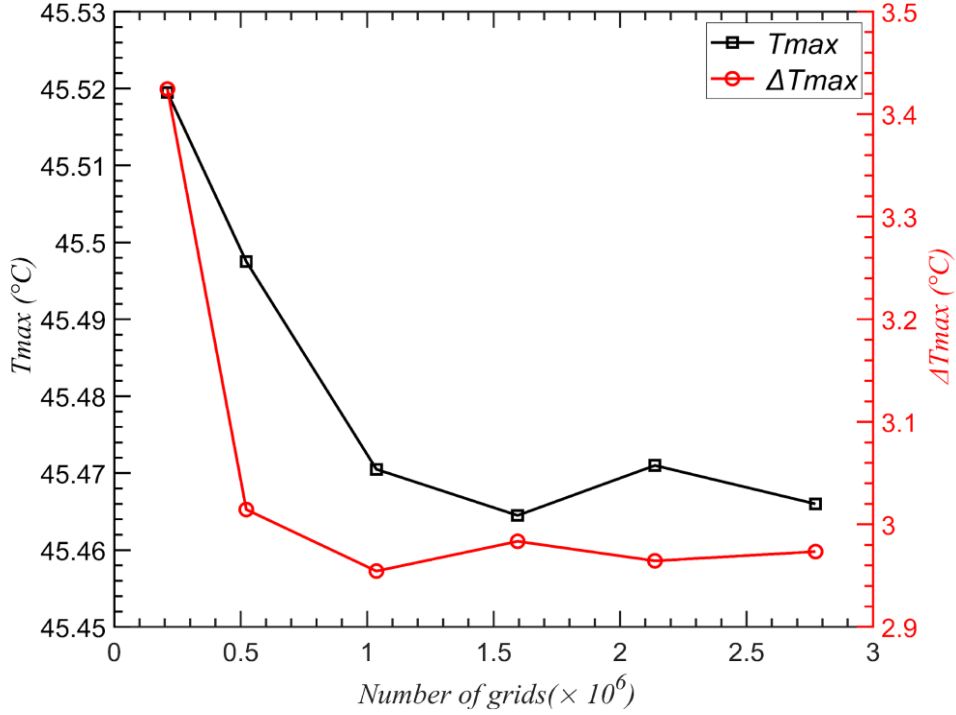

FIG. 5: Convergence analysis: Variation of results ($T_{max}$ and $\Delta T_{max}$) with increasing grid density

## IV. RESULTS AND DISCUSSION

The research aims to minimize the $T_{max}$ and $\Delta T_{max}$ of a battery by investigating the influence of factors such as inlet air velocity, tapered inlet air manifold, and the number of secondary outlets on the cooling performance. The optimization model considers lithium as the battery material, with operational parameters set at 25 °C ambient temperature, 2.5C discharge rate, and 3 m/s inlet air velocity. The objective is to improve the cooling system to enhance the battery's thermal management and ensure safe and effective operation, focusing on understanding how these aspects affect cooling performance.

### A. Effect of Inlet Air Velocity

The air inlet velocity is increased from 3 to 5 m/s with a step size of 0.5 m/s while keeping the remaining boundary conditions unchanged to replicate the thermal state of the battery pack under different airflow velocities. Fig. 6 depicts the quantitative relationship between $T_{max}$, $\Delta T_{max}$, and inlet air velocity. A decreasing trend is observed in $T_{max}$ as the air inlet velocity increases. The lowest temperature of the battery pack (40.9°C) is reached at an air inlet velocity of 4.5 m/s, after which it increases again. In contrast, $\Delta T_{max}$ initially shows an increasing trend up to 3.5 m/s, followed by a decrease until 4.5 m/s. Beyond 4.5 m/s, $\Delta T_{max}$ starts increasing once again.

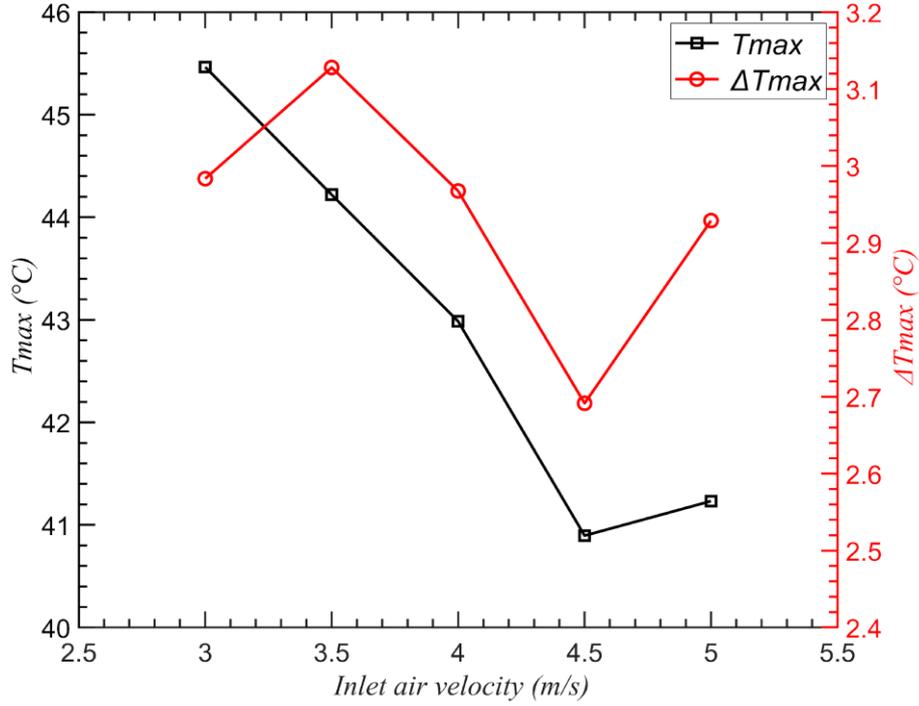

FIG. 6: Relationship between air inlet velocity and thermal dynamics, specifically focusing on maximum temperature ($T_{max}$) and temperature fluctuation ($\Delta T_{max}$).

According to Fig. 6, the battery pack achieves the lowest values of $T_{max}$ and $\Delta T_{max}$ at 40.9 °C and 2.6915 °C, respectively, when the air velocity setting for the intake is 4.5 m/s. Therefore, the baseline condition with a 4.5 m/s inlet velocity is considered for further optimization. Fig. 7(a) displays the battery cell temperatures at various intake velocities, revealing a decrease in temperature for each cell when the velocity rises from 3 m/s to 4.5 m/s. In Fig. 7(b), the velocity in each channel at different air inlet velocities is observed and measured at the center of the channel. Furthermore, Fig. 8 illustrates the simulated three-dimensional temperature and velocity distributions within the battery module for an inlet air velocity of 3 m/s. In comparison, Fig. 9 depicts the corresponding distributions for an inlet air velocity of 4.5 m/s. The findings indicate that the inlet air velocity significantly influences the cooling performance of the BTMS. Varying the inlet air velocity from 3 to 4.5 m/s decreases $T_{max}$ from 45.47 °C to 40.9 °C, representing a temperature decrease of 4.57 °C (10.05%). Similarly, $\Delta T_{max}$ decreases from 2.98 °C to 2.69 °C, corresponding to a reduction of 0.29 °C (9.79%). These results highlight the importance of optimizing the air inlet velocity to achieve improved cooling efficiency in the BTMS.

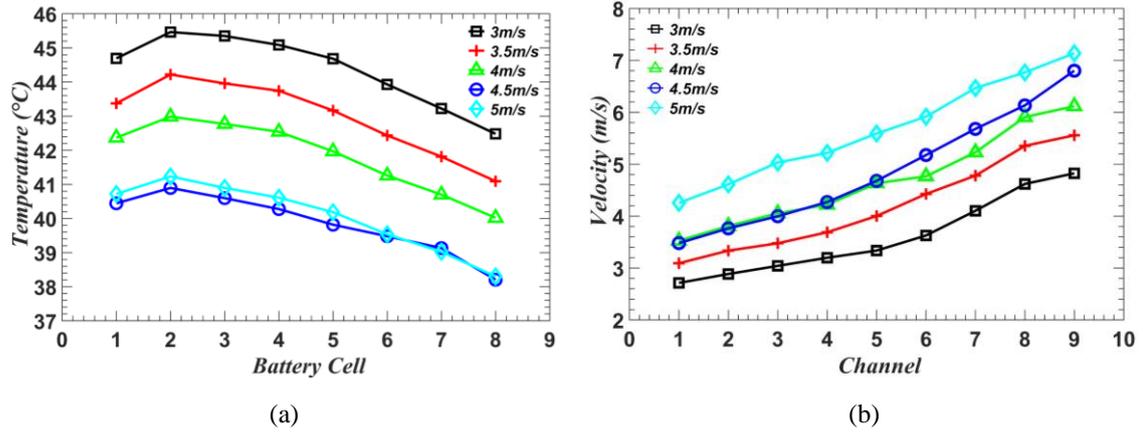

FIG. 7: (a) Temperature distribution across battery cells as measured during operation (b) Velocity profiles within channels depicted at varying airflow rates

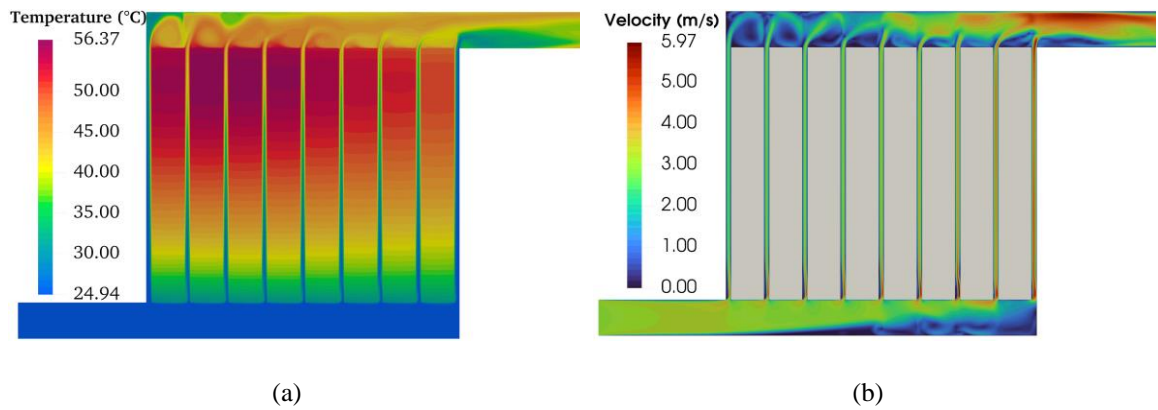

FIG. 8: (a) Temperature distribution across the domain at steady-state conditions. (b) Contour plot illustrating the velocity profile at the air inlet with a velocity of 3 m/s

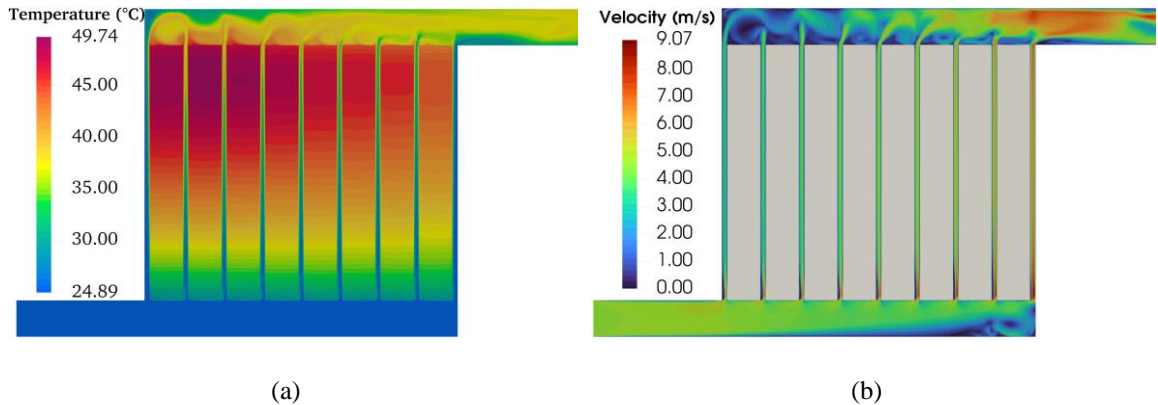

FIG. 9: (a) Temperature distribution across the domain at steady-state conditions. (b) Contour plot illustrating the velocity profile at the air inlet with a velocity of 4.5 m/s

## B. Effect of Tapered Inlet Manifold

The present investigation focuses on studying the impact of a tapered inlet manifold on the battery's temperature distribution. The air-outlet manifold's height is constant, and the cooling channel's width remains at its initial value of d = 3 mm. The air-inlet velocity is maintained at 4.5 m/s. Several air-inlet angles (3 mm, 5 mm, 8 mm, 9 mm, 10 mm, 15 mm, and 20 mm) are considered to examine this effect. Numerical simulations are performed to determine the $T_{max}$ and $\Delta T_{max}$ under each angle.

Fig. 10 demonstrates how the air-inlet angle affects $T_{max}$ and $\Delta T_{max}$. $T_{max}$ initially exhibits an upward trend as the air-inlet angle increases, followed by a gradual drop and stabilization. Similarly, $\Delta T_{max}$ shows an initial increase followed by stabilization. Notably, the cases with A = 9 mm or 10 mm show relatively smaller values for both evaluation indices, indicating improved thermal performance. Fig. 11 depicts the temperature and velocity contours inside the battery module for a tapered inlet manifold with A = 3 mm, while Fig. 12 and Fig. 13 illustrate the corresponding distributions for A = 9 mm and A = 20 mm, respectively. The temperature contours consistently reveal that the area closest to the battery's inlet side experiences the highest temperatures.

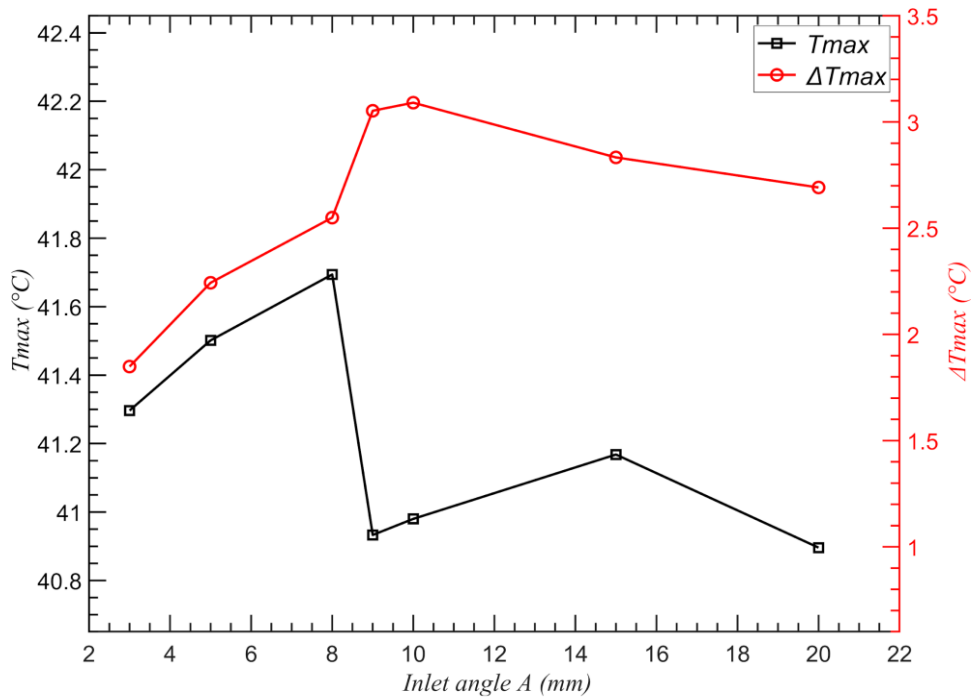

FIG. 10: Analysis of the variation of $T_{max}$ and $\Delta T_{max}$ at different heights A, measured at defined locations on the battery.

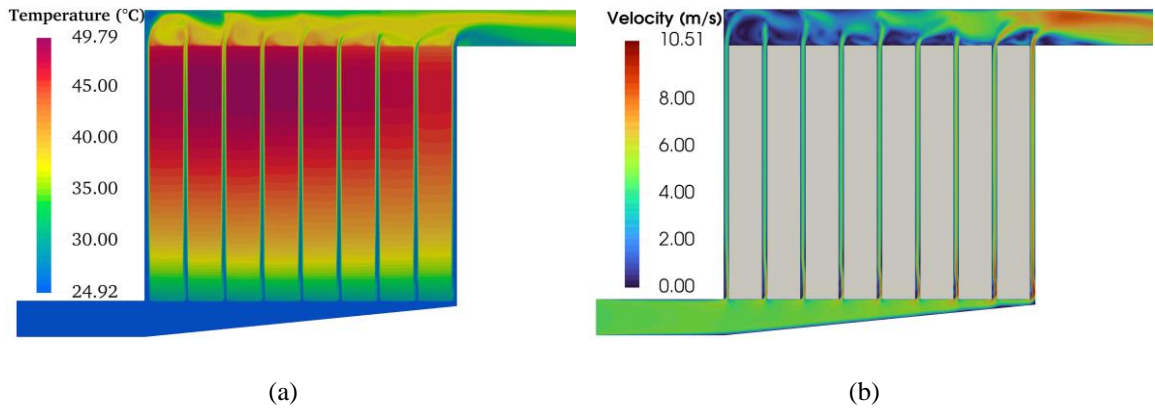

|(a)|(b)|

FIG. 11: (a) Temperature distribution across the tapered inlet manifold (A = 3mm), illustrating thermal variations along the length of the manifold (b) Contour plot showing velocity distribution within the tapered inlet manifold (A = 3mm), highlighting fluid flow patterns and velocity gradients within the manifold geometry

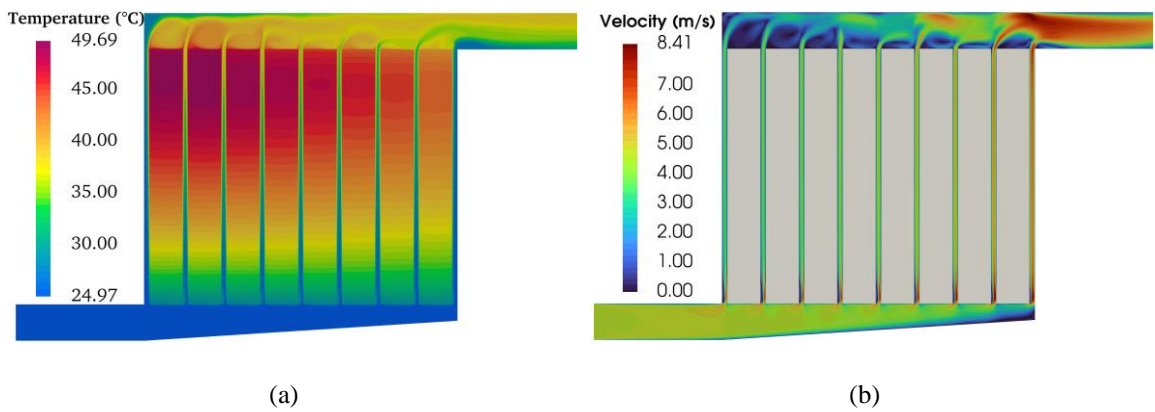

|(a)|(b)|

FIG. 12: (a) Temperature distribution across the tapered inlet manifold (A = 9mm), illustrating thermal variations along the length of the manifold (b) Contour plot showing velocity distribution within the tapered inlet manifold (A = 9mm), highlighting fluid flow patterns and velocity gradients within the manifold geometry

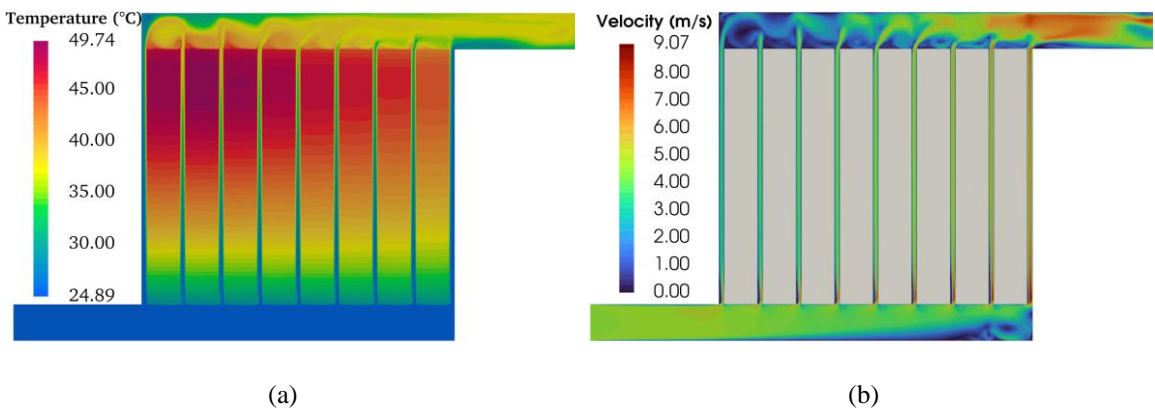

|(a)|(b)|

FIG. 13: (a) Temperature distribution across the tapered inlet manifold (A = 20mm), illustrating thermal variations along the length of the manifold (b) Contour plot showing velocity distribution within the tapered inlet manifold (A = 20mm), highlighting fluid flow patterns and velocity gradients within the manifold geometry

Fig. 14(a) illustrates the temperature distribution of individual battery cells. It is observed that when the tapered inlet manifold angle is set to 9 mm, cells (3-8) show a significant decrease in temperature compared to other configurations. Fig. 14(b) also presents the air flow velocities within the coolant passages, with channels 1-3 exhibiting higher velocities and channels 4-9 demonstrating lower velocities. When A is set to 9 mm, or an air-inlet manifold angle of 3.68°, this configuration facilitates a consistent airflow over the coolant channels. Consequently, by utilizing a tapered inlet angle of 9 mm, the variation in velocity between channel 9 and channel 1 decreases from 3.32 m/s to 2.50 m/s.

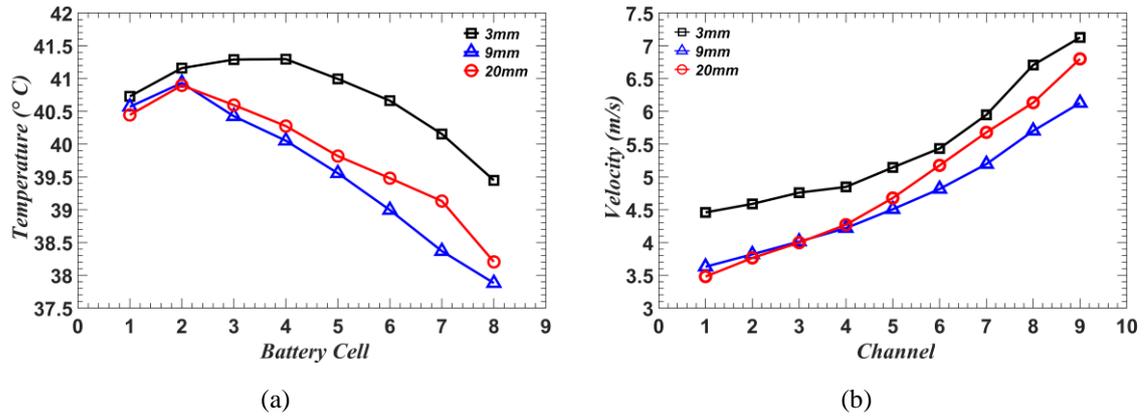

(a)            (b)

FIG. 14: (a) Temperature distribution across battery cells, highlighting thermal gradients (b) Velocity in channels at different heights: Illustrating flow dynamics at varied heights in the channel geometry

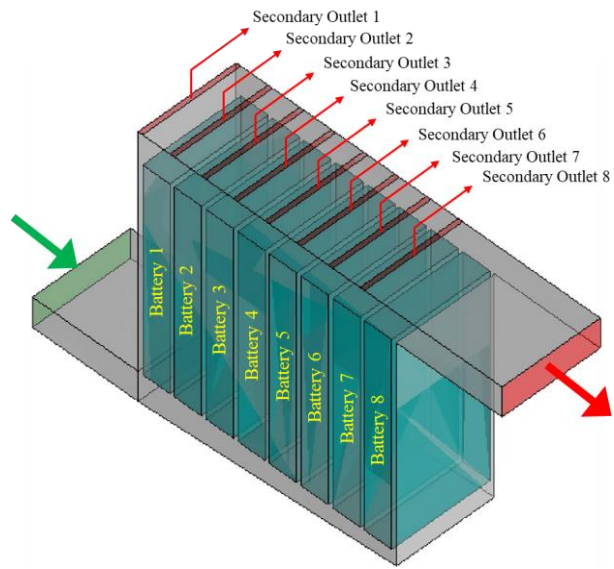

FIG. 15: Secondary outlet positions: Orientation of additional outlet channels additional to the primary outlet

## C. Effect of Secondary Outlet Number

The present study aims to assess how the number of secondary outlets influences the temperature distribution of the battery. The height of the air-inlet manifold (A) is adjusted to 9 mm, and the air-inlet velocity remains constant at 4.5 m/s. Fig. 15 displays the positioning of the secondary outlets, which have a width of 3 mm, corresponding to the distance between battery cells in the model. The secondary outlets are 65 mm long, the same length as a battery cell. On the battery module's top (outlet side), the secondary outlets are symmetrically positioned next to the cooling channel and labeled "Secondary Outlet-i" (i = 1, 2, ..., 7, 8). This setup allows for examining how the number and positioning of secondary outlets affect the battery's temperature distribution.

a. **Comparative Analysis of BTMS with and without Secondary Outlet**

In Fig. 16(a) and Fig. 16(b), the simulation results of two battery models are compared: one without any secondary outlets and another with the addition of Secondary Outlet-1. The temperature readings of each battery cell for the original model and the model with Secondary Outlet-1 are displayed in Fig. 16(a). The addition of a secondary outlet significantly affects $T_{max}$ and $\Delta T_{max}$, causing a noticeable temperature drop compared to the original model.

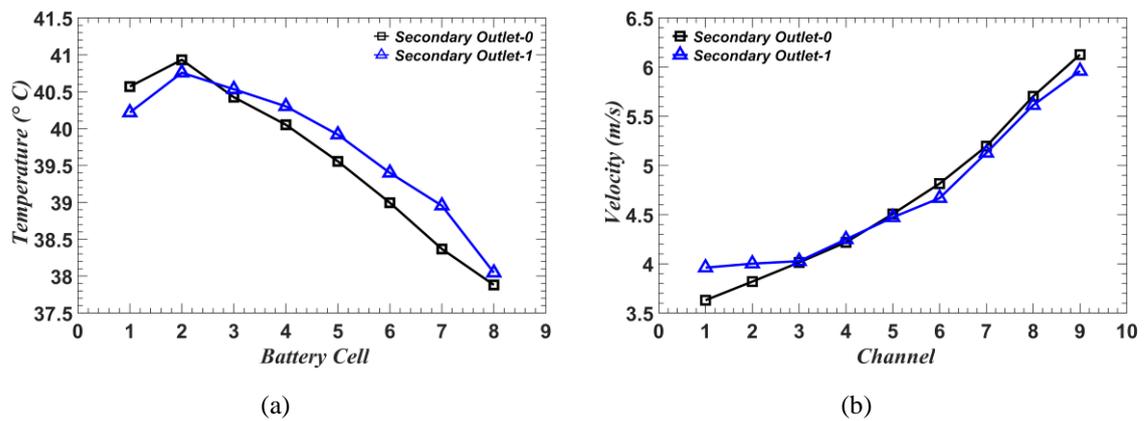

(a)        (b)

FIG. 16: (a) Temperature distribution across battery cells under operational conditions, illustrating variations in heat generation and dissipation within the battery pack. (b) Comparison of flow velocity profiles within BTMS channels featuring a single outlet (0 secondary outlet) versus enhanced flow dynamics with dual outlets (1 secondary outlet).

Fig. 16(b) shows the airflow velocities at the outlets of the original model and after adding Secondary Outlet-1. The original model exhibits a substantial velocity difference of up to 2.50 m/s between channel 9 and channel 1, resulting in a higher battery temperature near the intake than the outlet. However, with the addition of Secondary Outlet-1, the airflow velocities for cooling channels 1 to 4 increase, while those for cooling channels 5 to 9 decrease. This modification achieves an even distribution of airflow velocities within the cooling channels, reducing the maximum velocity difference to 2.0 m/s. These findings highlight the significant improvement in heat dissipation performance and the promotion of a more consistent temperature distribution within the BTMS.

### b. Comparative Analysis of the Number of Secondary Outlets

The study further investigates the impact of varying the number of secondary outlets on the cooling performance of the BTMS. Each secondary outlet is labeled "Secondary Outlet - N" (N = 1, 2, ..., 7, 8), representing the different outlet configurations. Fig. 17 displays the battery cell temperature, $T_{max}$, and $\Delta T_{max}$ for different numbers of secondary outlets. The results indicate a general downward trend in the battery temperature, $T_{max}$, and $\Delta T_{max}$ as the number of secondary outlets increases. Notably, when there are seven secondary outlets (Secondary Outlet - 7), the BTMS achieves its optimal cooling performance, with a reduction of 0.894 °C (2.18%) in $T_{max}$ and 2.22 °C (72.84%) in $\Delta T_{max}$ compared to the original model (with 0 secondary outlets).

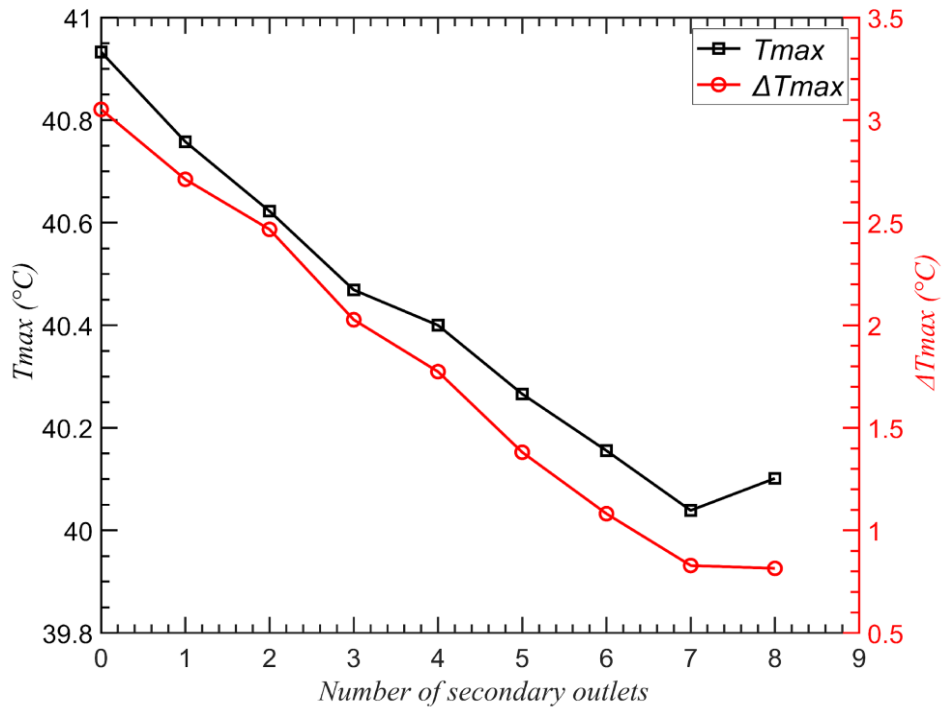

FIG. 17: Temporal variability analysis of maximum temperature ($T_{max}$) and its differential ($\Delta T_{max}$) across varying numbers of secondary outlets: Illustrating the fluctuations in $T_{max}$ and the differences ($\Delta T_{max}$) concerning the number of secondary outlets

The temperature and velocity distributions within the battery module are simulated for various secondary outlets (0, 1, 3, 5, and 7). The resulting three-dimensional distributions are observed in Fig. 18, Fig. 19, Fig. 20, Fig. 21, and Fig. 22, respectively. To further illustrate the effects of the secondary outlets, Fig. 23 displays the temperatures of each battery cell and the airflow velocities in each cooling channel for the original model (Secondary Outlet-0) and the model with Secondary Outlet-7. The airflow velocity comparison curves reveal an apparent increase in velocities for channels 1 to 5, while channels 6 to 9 experience a noticeable decrease. Similarly, the temperature comparison curves show decreased temperatures for batteries 1 to 4, while the temperature of batteries 5 to 8 increases. The observed changes are attributed to the reduction of pressure in the surrounding convergence plenum by the secondary outlets, resulting in a more uniform velocity distribution within the BTMS.

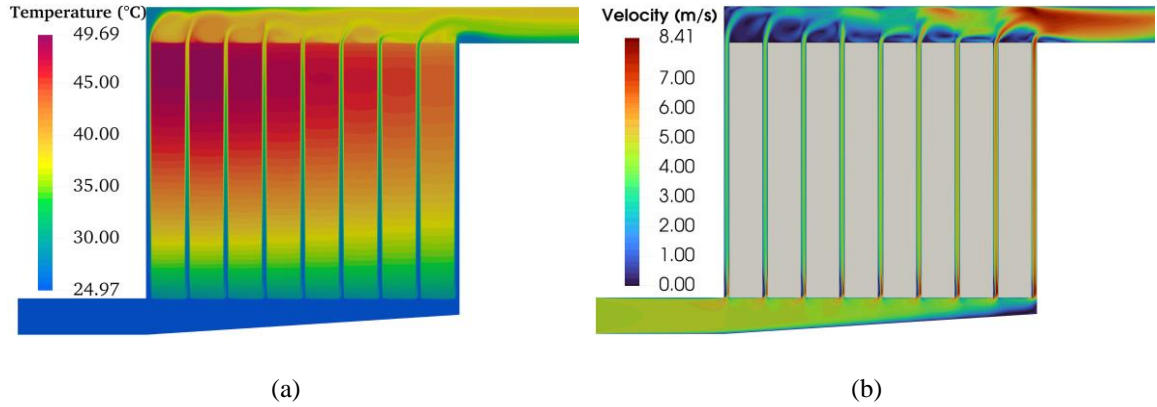

(a)　　　　　　　　　　　　　　　　　(b)

FIG. 18: (a) Temperature distribution across the battery module under specified operating conditions (b) Velocity contour of BTMS illustrating the flow characteristics, particularly highlighting the influence of the 0 secondary outlet, on the fluid dynamics within the system

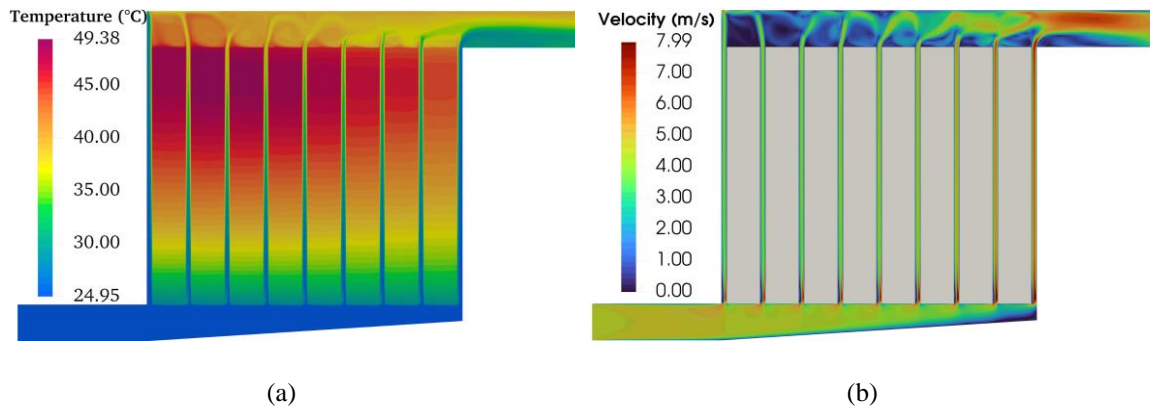

(a)　　　　　　　　　　　　　　　　　(b)

FIG. 19: (a) Temperature distribution across the battery module under specified operating conditions (b) Velocity contour of BTMS illustrating the flow characteristics, particularly highlighting the influence of the 1 secondary outlet on the fluid dynamics within the system

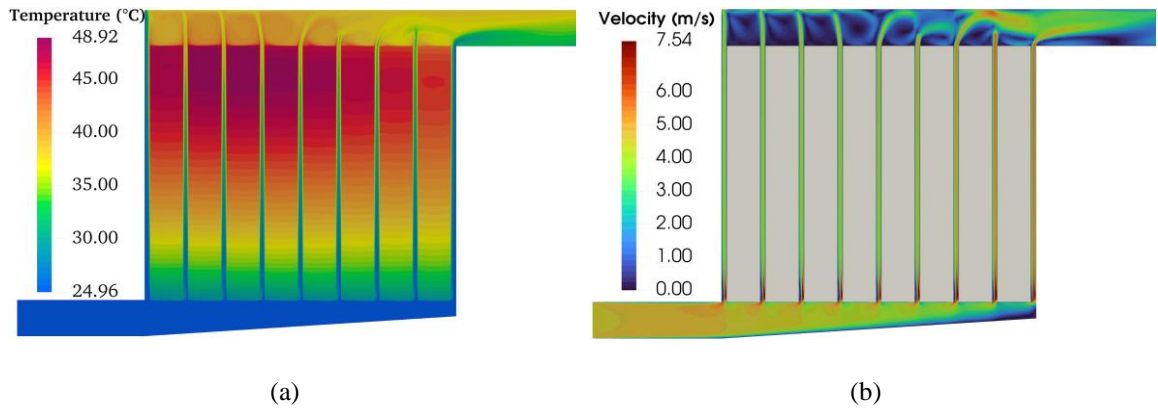

FIG. 20: (a) Temperature distribution across the battery module under specified operating conditions (b) Velocity contour of BTMS illustrating the flow characteristics, particularly highlighting the influence of the 3 secondary outlets on the fluid dynamics within the system

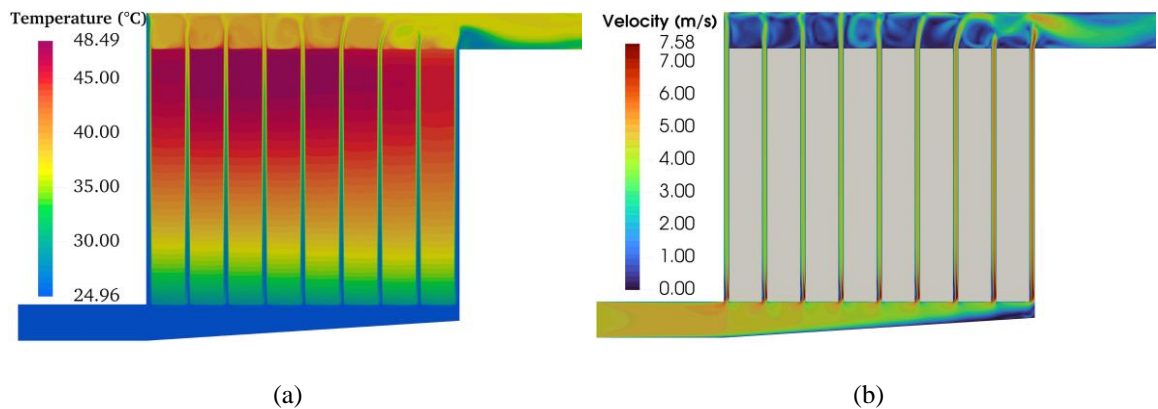

FIG. 21: (a) Temperature distribution across the battery module under specified operating conditions (b) Velocity contour of BTMS illustrating the flow characteristics, particularly highlighting the influence of the 5 secondary outlets on the fluid dynamics within the system

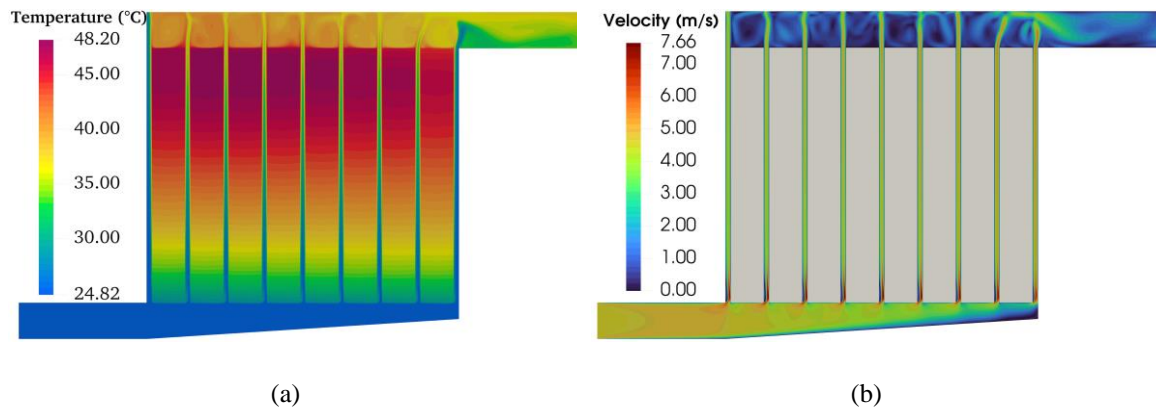

FIG. 22: (a) Temperature distribution across the battery module under specified operating conditions (b) Velocity contour of BTMS illustrating the flow characteristics, particularly highlighting the influence of the 7 secondary outlets on the fluid dynamics within the system

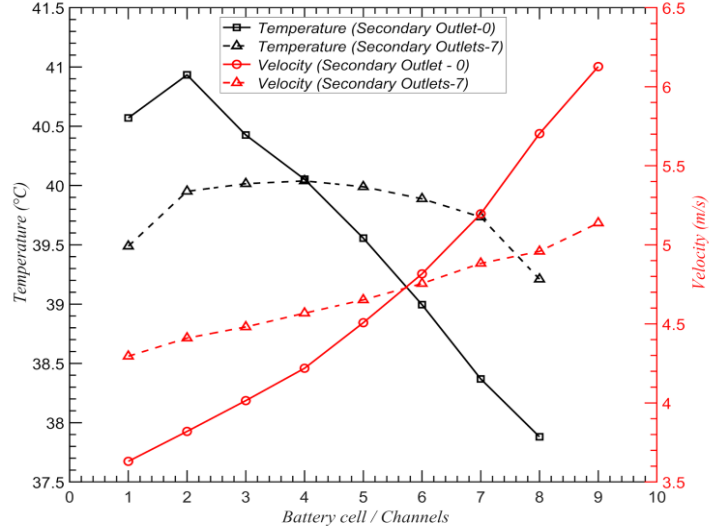

FIG. 23: Comparison of airflow velocity (m/s) and battery cell temperature (°C) between configurations with 0 and 7 secondary outlets, illustrating the impact of secondary outlet count on airflow distribution and battery thermal management.

The study underscores the significant impact of various design modifications on the thermal cooling performance of the Z-type BTMS. By optimizing parameters such as inlet air velocity, tapered inlet manifold, and the number of secondary outlets, notable improvements are achieved in temperature distribution, airflow uniformity, and heat dissipation ability. This holistic approach to structural enhancement substantially reduces $T_{max}$ and $\Delta T_{max}$, underscoring the efficacy of the Z-type BTMS design in enhancing battery pack cooling efficiency and overall thermal stability.

## V. CONCLUSION

The primary focus of this study is to optimize the structure of the FACS used for dissipating heat in LIBs. The study examines three key factors to enhance thermal performance: the inlet air velocity, the tapered inlet manifold, and the number of secondary outlets.

The flow field and temperature distribution within the battery module are thoroughly examined using CFD. Modifications are made to achieve a more uniform temperature and airflow rate in the BTMS by incorporating changes to the inlet air velocity, the tapered inlet manifold angle, and the number of secondary outlets. These optimizations significantly improve the heat dissipation capability of the BTMS and result in a more homogeneous temperature distribution. The study allows drawing the following conclusions:

1. The simulation findings are being compared with well-respected experimental data to confirm the validity of the CFD method. This comparison demonstrates the accuracy and effectiveness of the CFD approach in predicting the flow field and temperature distribution in the FACS for battery heat dissipation. A grid independence study is currently being performed to assess the impact of the grid resolution on the simulation results.

2. The inlet air velocity significantly impacts the thermal cooling performance of the BTMS. The study demonstrates that increasing the inlet air velocity from 3 to 4.5 m/s leads to notable improvements in temperature distribution. Specifically, the $T_{max}$ and $\Delta T_{max}$ decrease by 4.57 °C (10.05%) and 0.29 °C (9.79%), respectively, compared to the original 3 m/s inlet velocity.

3. Implementing a tapered inlet manifold in the BTMS significantly improves cooling performance. The study reveals that when the tapered inlet angle is set to 9mm, corresponding to a value of 3.68°, it results in remarkable reductions in the temperatures of battery cells 3-9.

    Additionally, the tapered inlet manifold is crucial in distributing the airflow more evenly in the cooling channels. Specifically, it reduces the difference between velocity in channel 9 and channel 1 from 3.32 m/s to 2.50 m/s, indicating a more uniform airflow along the coolant passages. This reduction in velocity difference is attributed to the optimized configuration of the tapered inlet manifold.

4. The number of secondary outlets significantly impacts the BTMS's heat dissipation ability. The investigation reveals that incorporating 7 secondary outlets improves the thermal cooling battery module performance. The results demonstrate that compared to the configuration having 0 secondary outlets, implementing 7 secondary outlets decreases $T_{max}$ by 0.894 °C (2.18%) and $\Delta T_{max}$ by 2.23 °C (72.84%).

5. The results demonstrate that the Z-type BTMS design reduces the battery pack's $T_{max}$ and $\Delta T_{max}$. Specifically, compared to the original model, the optimized Z-type BTMS configuration achieves a reduction of 5.43 °C (11.94%) in $T_{max}$ and 2.16 °C (72.2%) in $\Delta T_{max}$. These findings indicate a substantial enhancement in the battery pack's cooling performance and thermal stability.

Lowering $T_{max}$ and $\Delta T_{max}$ mitigates thermal stresses on battery cells, preventing capacity fade and accelerated aging, thus extending operational lifespan. Efficient thermal management improves overall energy efficiency by reducing internal resistance and mitigating energy losses, enhancing system performance. Consistent airflow distribution via the optimized BTMS configuration stabilizes thermal conditions, minimizing the risk of thermal runaway events and enhancing safety. These benefits translate to enhanced reliability, reduced maintenance, and extended EV battery lifespans, offering sustained performance and range to end-users.

## VI. SCOPE FOR FUTURE WORK

In this study, the focus is primarily on the number of secondary outlets. However, future work could explore other aspects of structural optimization for the BTMS, including investigating different secondary outlet widths, exploring combinations of outlet widths, optimizing using baffles or spoilers, and optimizing battery cell spacing. This study was limited to a BTMS battery pack containing only 8 single cells. However, actual battery packs consist of a much larger number of cells. Therefore, a promising direction for future work would be to investigate the performance and optimization of BTMS in large-scale battery packs.

In addition to the optimization method applied to prismatic batteries in this study, future work could extend the approach to other battery types, such as cylindrical cells (e.g., 18650) and pouch cells. By utilizing CFD, the focus could be on optimizing the physical structure design of these batteries. For 18650 cells, the spacing between each cell in the pack could be investigated, while for pouch cells, the arrangement of the cells within the pack could be explored. Furthermore, the optimization method employed in this study could be extended to investigate the applicability of other cooling systems, such as liquid cooling, heat pipes, and PCM cooling. For instance, future research on liquid cooling systems could focus on optimizing the physical structure design, including determining the optimal flow rate of the liquid and the spacing between each cell. Similarly, exploration could involve controlling the physical dimensions and the number of heat pipes for heat pipe cooling to enhance thermal performance.

## ACKNOWLEDGMENTS

The authors would like to acknowledge the National Supercomputing Mission (NSM) for providing the computational resources of 'PARAM Sanganak' at IIT Kanpur, which is implemented by C-DAC and supported by the Ministry of Electronics and Information Technology (MeitY) and Department of Science and Technology (DST), Government of India. The authors would also like to acknowledge the IIT-K Computer center (www.iitk.ac.in/cc) for providing the resources to perform the computation work.

## AUTHOR DECLARATIONS

### CONFLICT OF INTEREST

The authors have no conflicts to disclose.

### DATA AVAILABILITY

The data that support the findings of this study are available from the corresponding author upon reasonable request.